\documentclass[sigconf]{acmart}

\AtBeginDocument{%
  }

\copyrightyear{2024}
\acmYear{2024}
\setcopyright{acmlicensed}
\acmConference[CIKM '24]{Proceedings of the 33rd ACM International Conference on Information and Knowledge Management}{October 21--25, 2024}{Boise, ID, USA}
\acmBooktitle{Proceedings of the 33rd ACM International Conference on Information and Knowledge Management (CIKM '24), October 21--25, 2024, Boise, ID, USA}
\acmDOI{10.1145/3627673.3679782}
\acmISBN{979-8-4007-0436-9/24/10}

\usepackage{multirow}
\usepackage{balance}
\usepackage[ruled,linesnumbered]{algorithm2e}

\begin{document}

\title{DIIT: A Domain-Invariant Information Transfer Method for Industrial Cross-Domain Recommendation}

\author{Heyuan Huang}
\authornote{Both authors contributed equally to this research.}
\orcid{0009-0000-0320-3629}
\affiliation{%
  \institution{OPPO}
  \city{Shenzhen}
  \state{Guangdong}
  \country{China}
}
\email{huangheyuan2@oppo.com}

\author{Xingyu Lou}
\authornotemark[1]
\orcid{0009-0003-3180-0668}
\affiliation{%
  \institution{OPPO}
  \city{Shenzhen}
  \state{Guangdong}
  \country{China}
}
\email{louxingyu@oppo.com}

\author{Chaochao Chen}
\orcid{0000-0003-1419-964X}
\affiliation{%
  \department{College of Computer Science and Technology}
  \institution{Zhejiang University}
  \city{Hangzhou}
  \state{Zhejiang}
  \country{China}
}
\email{zjuccc@zju.edu.cn}

\author{Pengxiang Cheng}
\orcid{0009-0008-1218-3947}
\affiliation{%
  \institution{OPPO}
  \city{Shenzhen}
  \state{Guangdong}
  \country{China}
}
 \email{chengpengxiang@oppo.com}

\author{Yue Xin}
\orcid{0009-0003-5441-1403}
\affiliation{%
  \institution{OPPO}
  \city{Shenzhen}
  \state{Guangdong}
  \country{China}
}
 \email{xinyue@oppo.com}

\author{Chengwei He}
\orcid{0009-0008-6995-1580}
\affiliation{%
  \institution{OPPO}
  \city{Shenzhen}
  \state{Guangdong}
  \country{China}
}
\email{hechengwei@oppo.com}

\author{Xiang Liu}
\orcid{0009-0002-4360-8510}
\affiliation{%
  \institution{OPPO}
  \city{Shenzhen}
  \state{Guangdong}
  \country{China}
}
\email{liuxiang10@oppo.com}

\author{Jun Wang}
\authornote{Corresponding author.}
\orcid{0000-0002-0481-5341}
\affiliation{%
  \institution{OPPO}
  \city{Shenzhen}
  \state{Guangdong}
  \country{China}
}
\email{junwang.lu@gmail.com}

\renewcommand{\shortauthors}{Heyuan Huang et al.}

\begin{abstract}
  Cross-Domain Recommendation (CDR) have received widespread attention due to their ability to utilize rich information across domains. However, most existing CDR methods assume an ideal static condition that is not practical in industrial recommendation systems (RS). Therefore, simply applying existing CDR methods in the industrial RS environment may lead to low effectiveness and efficiency. To fill this gap, we propose DIIT, an end-to-end \textbf{D}omain-\textbf{I}nvariant \textbf{I}nformation \textbf{T}ransfer method for industrial cross-domain recommendation. Specifically, We first simulate the industrial RS environment that maintains respective models in multiple domains, each of them is trained in the incremental mode. Then, for improving the effectiveness, we design two extractors to fully extract domain-invariant information from the latest source domain models at the domain level and the representation level respectively. Finally, for improving the efficiency, we design a migrator to transfer the extracted information to the latest target domain model, which only need the target domain model for inference. Experiments conducted on one production dataset and two public datasets verify the effectiveness and efficiency of DIIT.
\end{abstract}



\maketitle

\section{Introduction}
\begin{figure}[!t]
    \centering
    \begin{tabular}{@{\extracolsep{\fill}}c@{}c@{\extracolsep{\fill}}}
            \includegraphics[scale=0.145]{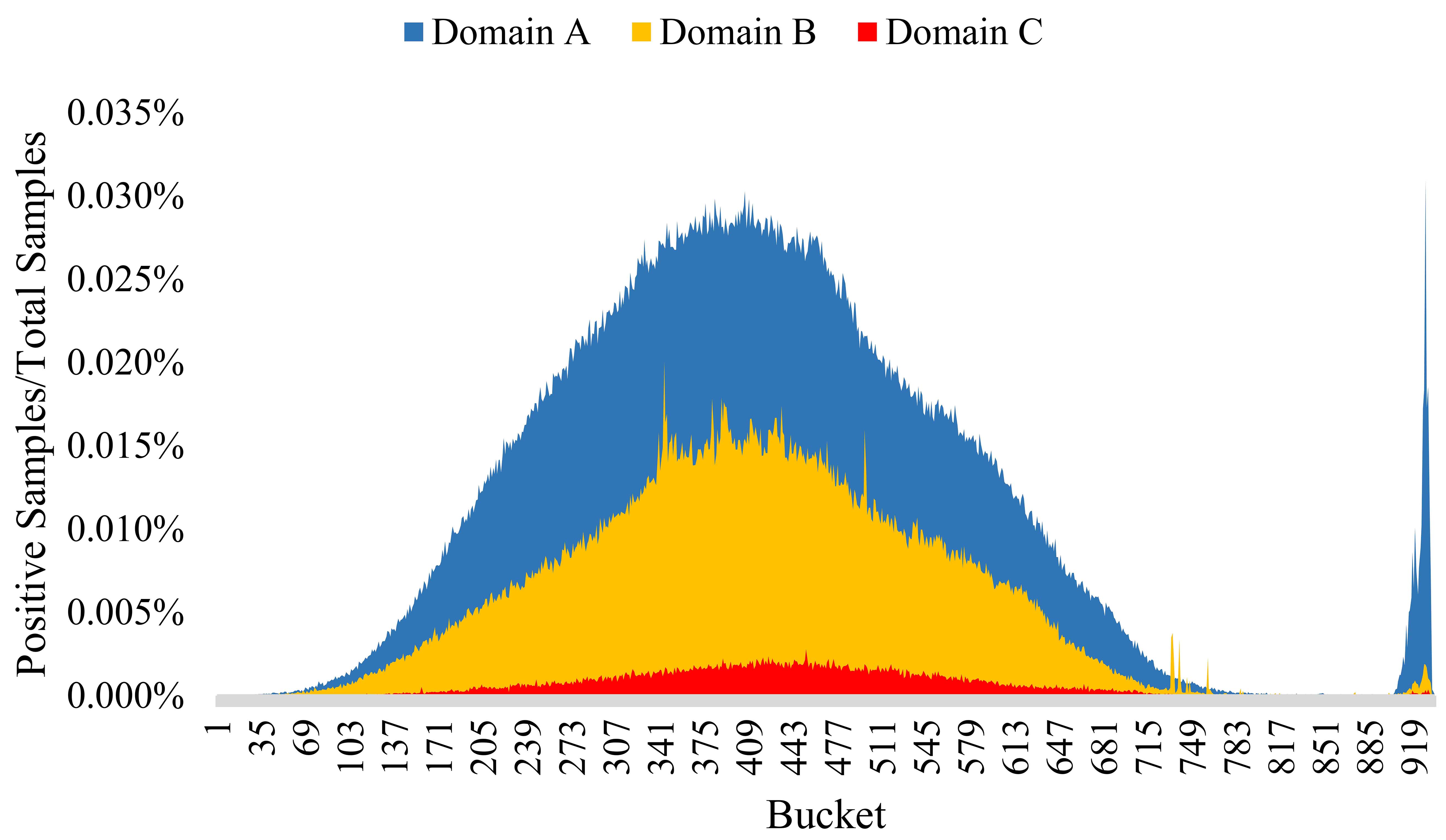} &
            \includegraphics[scale=0.145]{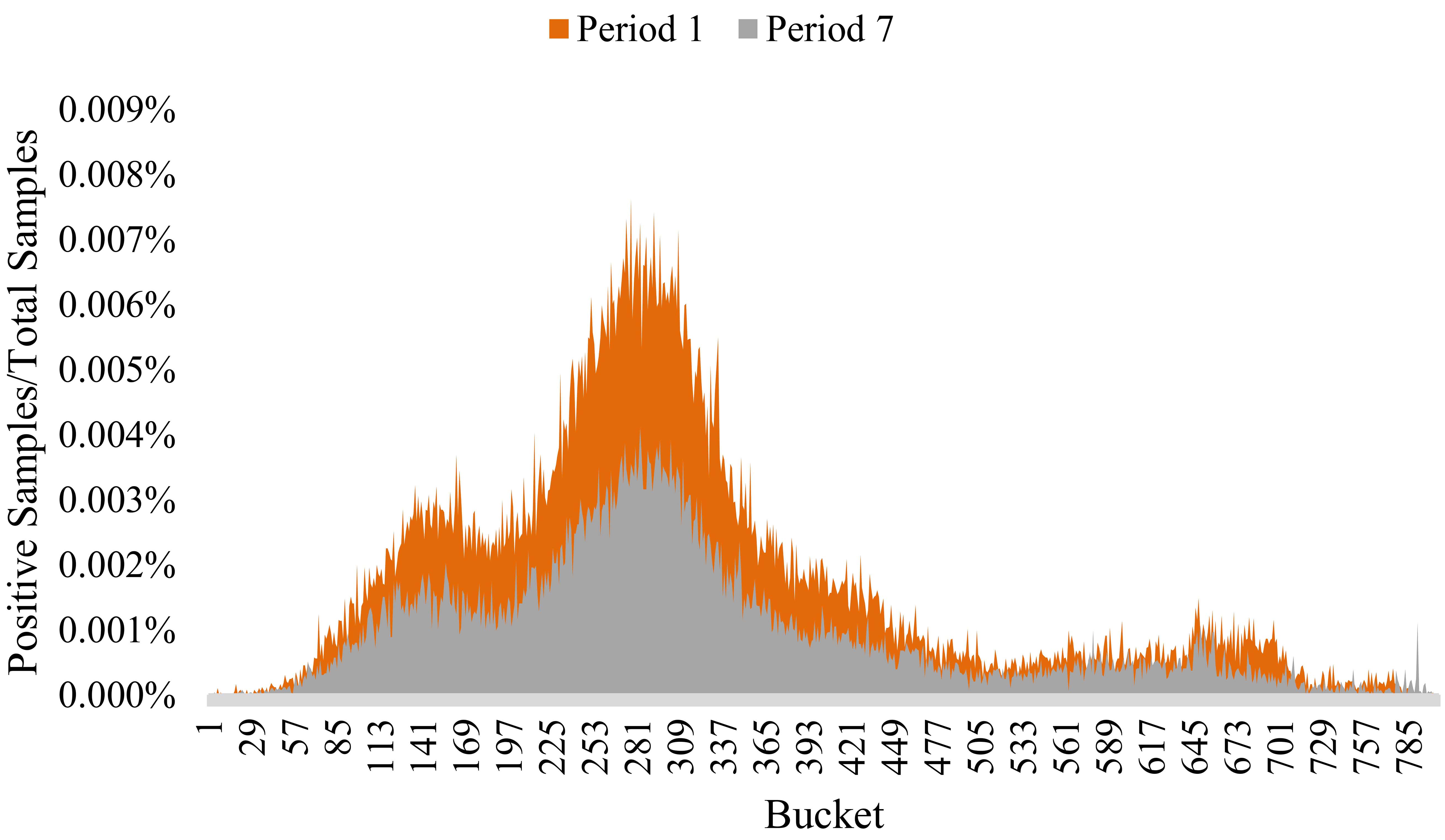}\\
            (a) & (b)\\
    \end{tabular}
    \caption{Visualization of the data distribution shift over domains (a) and time (b) using PCA. Area under curve represents CTR, and the data is collected from our production system.}
    \Description{Figure 1 shows the analysis of the data collected from OPPO's production system. By visualizing the data distribution in different domains at the same time and in the same domain at different times, we verified the existence of the data distribution shift over domains and time.}
    \label{fig:image_with_table1}
 \end{figure}

Large-scale commercial platforms typically contain multiple domains, and users are divided into many domains for different purposes, which leads to the data distribution shift over domains as shown in Figure \ref{fig:image_with_table1} (a). Therefore, Cross-Domain Recommendation (CDR) has emerged to transfer information across domains \cite{Zhu21,Cao23,Zang23}. Most existing CDR methods are based on an ideal static assumption that users' interests do not change much in a short period \cite{Zhang24,Zhang22,Liu23}. However, as shown in Figure \ref{fig:image_with_table1} (b), in the industrial RS environment, users' immediate interests are constantly changing, which leads to the data distribution shift over time \cite{Zeng24}. Therefore, it is indispensable to improve the efficiency of CDR methods to capture users’ interests immediately. 

In this paper, we focus on the problem of how to maximize the transmission of beneficial information across domains in the industrial RS environment. Nevertheless, our work faces the following two challenges:

\textbf{CH1}: \textit{How to improve the effectiveness of CDR methods in the industrial RS environment?} Since information in each domain can be divided into domain-specific and domain-invariant, where the former is only beneficial to the domain itself, the latter is beneficial to multiple domains \cite{Zhou23,Liu22}. Therefore, indiscriminately transferring information from the source domain to the target domain may be harmful. To tackle this challenge, most existing CDR methods assume the existence of overlapped samples between different domains, and regard these samples as "bridges" to transfer domain-invariant information across domains \cite{Zhao23,Cao22,Chen22}. However, in the industrial RS environment, it is difficult to fully obtain overlapped samples due to reasons like the large number of source domains involved and privacy protection, which seriously harm the effectiveness of these CDR methods.

\textbf{CH2}: \textit{How to improve the efficiency of CDR methods in the industrial RS environment?} Since large-scale commercial platforms contain multiple domains and each domain maintains its own model that trains in the incremental mode \cite{Wang20,Katsileros22,Ash20}, directly applying CDR methods to the industrial RS environment is sub-optimal. Recently, several works were proposed to specifically optimize CDR methods in the industrial RS environments and have achieved good results. However, we argue that these works fail to solve \textbf{CH2} well because they need additional computation and storage resources \cite{Zhang22} or need to keep the source domain model as external information to assist the inference in the target domain \cite{Liu23}, which leads to low efficiency.

To address the above challenges, in this paper, we propose DIIT, an end-to-end \textbf{D}omain-\textbf{I}nvariant \textbf{I}nformation \textbf{T}ransfer method for industrial cross-domain recommendation. For \textbf{CH1}, we design two extractors to fully extract the domain-invariant information. The first one named the domain-invariant information extractor at the domain level, which is composed of a gating network and uses the target domain model to adaptively guide the aggregation of multiple source domain models. The other one named the domain-invariant information extractor at the domain level, which is composed of an adversarial network and uses adversarial learning to align the distribution of representations output by source domain models and the target domain model respectively. Through these two extractors, we can better extract domain-invariant information. For \textbf{CH2}, we design a domain-invariant information migrator, which is composed of a multi-spot knowledge distillation (KD) network to transfer the extracted information to the latest target domain model. It is worth noting that, KD have a characteristic that not only allows a flexible development of teacher and student models with different architectures, but also requires only the student model in the inference phase. Therefore, we can transfer information from multiple source domain models with different structures, and keep only the target domain model for inference, which is more practical in the industrial RS environment and execute an efficient inference.

In summary, our contributions are as follows:
\begin{itemize}
    \item We first discuss an important but neglected research direction that how to perform efficient recommendations across domains in the industrial RS environment. As a potential solution, we propose DIIT, an end-to-end domain-invariant information transfer method for industrial cross-domain recommendation.
    \item We further analyze the challenges of how to improve the performance and efficiency of CDR methods in the industrial RS environment. For the former challenge, we design two extractors to extract domain-invariant information between the source domains and the target domain at two levels respectively. For the latter challenge, we design a migrator to transfer domain-invariant information from the source domain models to the target domain model, while only keeping the target domain model for inference.
    \item We finally conduct extensive experiments on three datasets of different magnitudes, including one production dataset and two public datasets, to verify the effectiveness and efficiency of DIIT.
\end{itemize}

\section{Related Works}
\label{sec:headings}

\subsection{Cross-Domain Recommendation in Incremental Learning}
Cross-Domain Recommendation (CDR) is widely used to transfer domain-invariant information across domains \cite{Zhu21,Cao23,Zang23}. Specifically, CDR methods divide different domains into source and target domains \cite{Zhu23}, and on the one hand, obtain the domain-invariant information from the source domain to improve the effectiveness of the model in the target domain. On the other hand, preserves as much domain-specific information of the target domain as possible \cite{Chen24,Ouyang20,Gan24,Zhang232}.

However, most of the above methods cannot be well adapted to the industrial RS environment \cite{Wang20,Katsileros22,Ash20}. InMSR \cite{Zhang24} is a recent multi-domain incremental method that combines information from domain, time, and time-domain dimensions respectively. KEEP \cite{Zhang22} and CTNet \cite{Liu23} involve the cross-domain recommendation of incremental learning, which is closest to the research direction in this paper. Among them, KEEP is a two-stage industrial knowledge extraction and plugging framework, but still suffers from disadvantages such as the need for additional computation and storage resources, and the need for a synchronization strategy to ensure consistency between the knowledge extraction and inference. CTNet is a lightweight method that transfers information from the time-evolving source domain to the time-evolving target domain. CTNet first extends the target domain model to be a two-tower architecture, including a source tower and a target tower, which are consistent with the corresponding source/target domain model respectively. Then, design an adapter to transfer information from the source tower to the target tower both in the training and inference phases, which leads to lower efficiency. Furthermore, we argue that the methods above cannot fully extract the domain-invariant information through simple addition or mapping layers. As a potential solution, in our paper, we proposed DIIT, which can efficiently transfer domain-invariant information in the industrial RS environment.

\subsection{Adversarial Learning}
Adversarial learning is a research direction of transfer learning, which can extract domain-invariant information by aligning domains \cite{Huang11,He18,Dayal23,Elgammal17}. Its core is to design a discriminator to distinguish which domain the sample comes from. By confusing the discriminator, it can obtain domain-invariant information across domains. Adversarial learning works because it is equivalent to minimizing the Jensen-Shannon divergence between different distributions \cite{Goodfellow20}. DANN \cite{Ganin16} is an early work that uses adversarial learning to align the labeled source domain and unlabeled target domain, thereby classifying target domain data with the help of the source domain. In recommendation, adversarial learning has also received widespread attention \cite{Su22,Hao21,Tan24,Zhang233}. For example, su et al.\cite{Su22} proposed an adversarial learning-based framework to train the target model together with a pre-trained source model. In this paper, to transfer domain-invariant information across domains in a fine-grained manner, we design an adversarial learning-based extractor to learn domain-invariant information by aligning the distribution of representations output by the source domain models and the target domain model respectively.

\begin{figure*}[!t]
  \centering
  \includegraphics[scale=0.053]{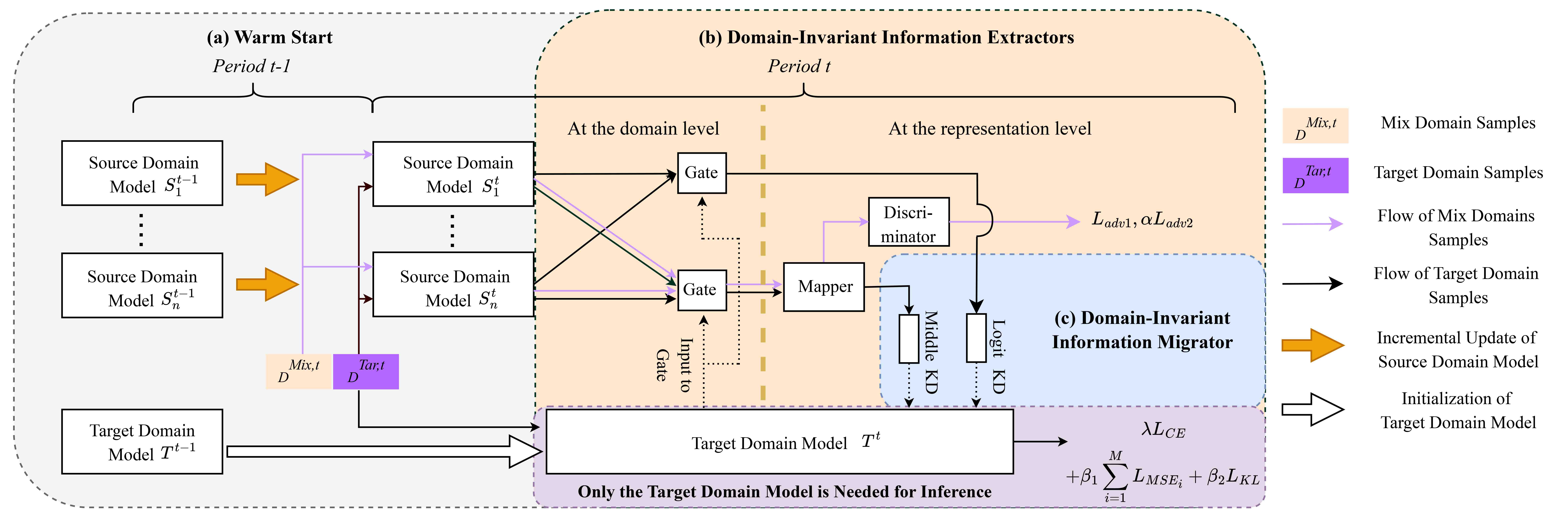}
  \caption{Model Architecture of the proposed DIIT. DIIT consists of three parts: A warm start module, a domain-invariant information extractor module and a domain-invariant information migrator module. Note that only the target domain model in the bottom right region is used for inference. The solid arrows in violet and black represent the sample flow of the mix dataset and the target domain dataset respectively. The larger arrows in orange and white represent incremental update of the source domain models and initialization of the target domain model respectively.}
  \Description{Figure 2 shows the overall framework of the proposed DIIT. Specifically, different color partitions and different arrows are used to illustrate its main modules, incremental update logic, and data flow.}
  \label{fig:model_architecture2}
\end{figure*}

\subsection{Knowledge Distillation}
Knowledge distillation (KD) is a model compression method based on the teacher-student learning framework \cite{Ojha23,Chen23} that was first proposed by Hinton et al. \cite{Hinton15}. In addition to model compression, KD can also be used to train a unified model to represent multiple models \cite{Zhu20}. When these models are come from different domains, information is transferred across domains. Existing KD methods can be divided into one-spot and multi-spot distillation \cite{Tian20,Song22}, where one-spot KD methods only transfer information from one layer of the teacher model, usually the logit layer. However, different layers of the teacher model have different semantic information, so multi-spot distillation is proposed, which provides more supervision signals for the student model by transferring the multi-layer information from the teacher network. In this paper, we use multi-spot KD to transfer information from multiple source domain models to the target domain model.

\section{Methodology}
\label{sec:Methodology}

\subsection{Problem Formulation}
\label{section Problem Formulation}
Most existing CDR methods assume an ideal static assumption that users' interests do not change much in a short period \cite{Zhang24,Zhang22,Liu23}. Therefore, simply applying them to the industrial RS environment will result in low effectiveness and efficiency. In this paper, we consider a cross-domain recommendation task in the industrial RS environment that contains multiple domains, each domain maintains its own model and is trained in the incremental mode. On this basis, our goal is to efficiently extract and transfer the domain-invariant information, thereby ultimately improve the recommendation effectiveness in the target domain. Formally, we take the samples from the target domains in the period \textit{t} as $D^{Tar,t} = (x^{Tar,t}, y)$, and collect source domain samples and target domain samples at a ratio of 1:1 to construct a mixed dataset in the period \textit{t} as $D^{Mix,t} = (x^{Mix,t}, d)$, where the size of $D^{mix,t}$ is consistent with the size of the target domain dataset $D^{Tar,t}$. $x^{Tar,t}$ and $x^{Mix,t}$ represent samples from the above two datasets respectively. $y \in \{ 0,1 \}$ is the label that indicates whether the sample was clicked. $t$ represents the period when the sample was collected. $d \in \{ 0,1 \}$ is the domain indicator that indicates whether the sample is from the target domain. Furthermore, we use $S_n^t(\cdot)$ to represent the model in each source domain in the period \textit{t} respectively, and use $T^t(\cdot)$ to represent the model in the target domain in the period \textit{t}, where $n \in \{ 1,...,N \}$ and $N$ represents the number of source domains. Simply speaking, we want to transfer as much domain-invariant information as possible from source domain models $S_n^t(\cdot)$ to the target domain model $T^t(\cdot)$ while preserving the domain-specific information of the target domain in the period \textit{t} of incremental learning. 

\subsection{Overall Framework}
The main framework of DIIT is shown in Figure \ref{fig:model_architecture2}, which mainly consists of three modules: a warm start module, a domain-invariant information extractor module and a domain-invariant information migrator module. The work process can be split into training and inference phases:
\begin{itemize}
\item
\textbf{Training:} Firstly, we design a warm start module to simulate the industrial RS environment, which maintains a unique model in each domain and is trained in the incremental mode independently (see part (a) of Figure \ref{fig:model_architecture2}). Secondly, we design a domain-invariant information extractor module that consists of two extractors to extract the domain-invariant information at the domain level and the representation level respectively (see part (b) of Figure \ref{fig:model_architecture2}). Specifically, the first extractor uses the target domain model to guide the aggregation of multiple source domain models adaptively through a gating network. The second extractor uses an adversarial network to align the distributions of the aggregated source domain representations and the target domain representations. Thirdly, we design a domain-invariant information migrator, which is composed of a multi-spot knowledge distillation (KD) network to transfer the extracted domain-invariant information from the source domain models to the target domain model robustly (see part (c) of Figure \ref{fig:model_architecture2}). Finally, we optimize the above modules synchronously with the recommendation task in the target domain, thereby efficiently utilizing both the domain-specific information from the target domain and the domain-invariant information from the source domains to improve the recommendation effectiveness in the target domain.
\end{itemize}

\begin{itemize}
\item
\textbf{Inference:} After training, the domain-invariant information from the source domains has actually been fully learned by the target domain model. And due to the characteristic of KD that requires only the student model for inference, we only need to keep the target domain model during the inference phase, which undoubtedly improves the efficiency of DIIT.
\end{itemize}

\subsection{Warm Start}
First of all, to simulate the industrial RS environment, each domain maintains an independent model that is trained in the incremental mode to capture domain-specific information. Taking the period \textit{t} as an example, we first learn the period \textit{t}'s model of each source domain independently using the latest incoming data, which is shown by the orange arrows in part (a) of Figure \ref{fig:model_architecture2}. For the target domain model, we initialize the period \textit{t}'s model using the period \textit{t-1}'s model in a warm start manner, which as shown by the white arrow in part (a) of Figure \ref{fig:model_architecture2}. As mentioned in section \ref{section Problem Formulation}, in the cross-domain recommendation task we consider, the structure of each source domain model can be different, and it is more practical in the industrial RS environment. In addition, since DIIT is plug-and-play, assuming that we plug the proposed DIIT for the first time in period \textit{t}, there is no need to train the target domain model from scratch using all available data, which is undoubtedly efficient.

\subsection{Domain-invariant Information Extractors}
\label{section Domain-invariant Information Extractors}
As mentioned above, most existing CDR methods require large amounts of source domain data, which is impractical in the industrial RS environment \cite{Zhao23,Cao22}. To handle this problem and fully mine the domain-invariant information, as shown in part (b) of Figure \ref{fig:model_architecture2}, we design two extractors to extract the domain-invariant information from the source domain model at the domain level and the representation level respectively. Our method requires only the source domain models and a small number of source domain samples for training, we will introduce them in the rest of this section.

\subsubsection{at the domain level} To extract the domain-invariant information at the domain level while ensuring the training time does not significantly increase as the number of source domains increases, we design the first extractor to aggregate the representations and the category probabilities (i.e. logits) output by multiple source domain models. Strategically, we input the period \textit{t}'s target domain samples into the latest source and target domain models respectively. Next, through a gating network, the target domain model representations are used to adaptively guide the aggregation of representations and logits of each source domain model. Specifically, the output of the domain-invariant information extractor at the domain level is formulated as:
\begin{equation}
\mathbf{e_s^t} = \sum\limits_{n=1}^{N}g_{s_n}^t\mathbf{e}_{s_n}^t,\quad\quad \mathbf{Z_s^t}=\sum\limits_{n=1}^{N}g_{s_n}^t\mathbf{Z}_{s_n}^t,
\label{eq1}
\end{equation}
where $\mathbf{e}_{s_n}^t,\mathbf{Z}_{s_n}^t = S_n^t(x^{Tar,t})$ represents the representation and the logit output by the $n$-th source domain model in the period \textit{t}. $g_{s_n}^t$ represents the output of the gating network,  which is composed of a two-layer multi-layer perceptron (MLP), and uses the softmax function as activation:
\begin{equation}
g_{s_n}^t = softmax(W_{s_n}^t(\mathbf{e}_{T}^t)),
\end{equation}
where $W_{s_n}^t$ represents the transformation matrix of the gating network, $\mathbf{e}_{T}^t = T^t(x^{Tar,t})$ represents the representation output by the target domain model in the period \textit{t}.

\subsubsection{at the representation level}
Despite obtaining $\mathbf{e}_s^t$ that contains the domain-invariant information, we argue that it is still necessary to obtain more invariant information from a fine-grained perspective. Therefore, we design the second extractor to extract the domain-invariant information from multiple source domain models at the representation level. Strategically, we align the representation distributions of the source domains and the target domain through an adversarial network, which can separate the domain-invariant information and the domain-specific information.  

As shown in part (b) of Figure \ref{fig:model_architecture2}, the second extractor consists of two parts, a mapper and a discriminator. As shown by the blue arrow in the figure, we input the representations of samples from different domains into the mapper and the discriminator successively. Note that the purpose of the discriminator is to determine whether the sample comes from the target domain, while the purpose of the discriminator is to confuse it. This is actually a min-max problem, and when the discriminator cannot correctly determine the source of the representations, it is considered that the distributions of the source domain representation and the target domain representation have been aligned. Specifically, we feed samples from the mixed dataset $D^{Mix,t}$ into each source domain model separately, and use the first extractor to obtain the aggregated representations $\mathbf{e}_{adv}^t$. Next, we feed $\mathbf{e}_{adv}^t$ into the mapper, which consists of a simple linear transformation:
\begin{equation}
\mathbf{e}_{adv}^t = W_{mapper}^t(\mathbf{e}_{adv}^t),
\label{eq3}
\end{equation}
where $W_{mapper}^t$ represents the transformation matrix of the mapper. Next, we feed $\mathbf{e}_{adv}^t$ into the discriminator to distinguish whether $\mathbf{e}_{adv}^t$ comes from the source domains or the target domain. The discriminator consists of a two-layer MLP, and uses the sigmoid function as activation.
\begin{equation}
\hat{d} = sigmoid(W_{dis}^t(\mathbf{e}_{adv}^t)),
\end{equation}
where $W_{dis}^t$ represents the transformation matrix of the discriminator. After getting the predicted $\hat{d}$, Our goal is to minimize the cross-entropy loss as:
\begin{equation}
L_{adv1}(\mathbf{\Theta}_{dis}) = -\frac{1}{|D^{Mix,t}|}\sum\limits_{i=1}^{|D^{Mix,t}|}d_ilog\hat{d}_i-(1-d_i)log(1-\hat{d}_i),
\label{eq5}
\end{equation}
where $\mathbf{\Theta}_{dis}^t$ represents the parameters of the discriminator. Note that the discriminator and other parts of the model will be trained successively in one epoch to achieve an effect similar to the gradient inversion. That is, during the optimization of $L_{adv1}$, all parameters will be frozen except that of the discriminator. After optimizing the discriminator, what we have to do is to confuse it through the mapper. Specifically, we maximize a loss function $L_{adv2}$ while freeze parameters of the discriminator during the optimization:
\begin{equation}
L_{adv2}(\mathbf{\Theta}_{mapper}^t) = -\frac{1}{|D^{Mix,t}|}\sum\limits_{i=1}^{|D^{Mix,t}|}d_ilog\hat{d}_i-(1-d_i)log(1-\hat{d}_i),
\label{eq6}
\end{equation}
where $\mathbf{\Theta}_{mapper}^t$ represents the parameters of the mapper. The optimization of $L_{adv2}$ will be performed together with other losses of DIIT, we will introduce it in detail in section \ref{Optimization and Inference}. Next, we pass $\mathbf{e}_{s}^t$ obtained from Eq. (\ref{eq1}) through the mapper, so that it can obtain the domain-invariant information at the representation level.

\subsection{Domain-invariant Information Migrator}
Once the representations with domain-invariant information are obtained, how to transfer them is the next challenge. In order to ensure efficient inferring and improve the generalization of the target domain model, we design a multi-spot KD network to transfer domain-invariant information efficiently from the source domain models to the target domain model, which is shown in part (c) of Figure \ref{fig:model_architecture2}. Strategically, to fully obtain multi-level information, we adopt the middle layer distillation and the logit layer distillation, which will be introduced in the rest of this section.

\subsubsection{middle layer distillation} We first let the middle layer representations of the target domain model imitate the corresponding representations of the source domain models, thereby accepting information across domains robustly. specifically, we use the Mean Square Error (MSE) loss to minimize the distribution gap between $\mathbf{e}_{s}^t$  output by the domain-invariant information extractor at the representation level and the corresponding representation $\mathbf{e}_T^t$ output by the target domain model:
\begin{equation}
    L_{MSE} = \frac{1}{|D^{Tar,t}|}\sum\limits_{i=1}^{|D^{Tar,t}|}(W_{KD}^t\mathbf{e_s}^t-\mathbf{e}_T^t)^2
\end{equation}
where $\mathbf{e}_T^t = f_{T}(x^{Tar,t})$ represents the representation output by the target model. $W_{KD}^t$ represents a transformation matrix in case when $\mathbf{e}_s^t$ and $\mathbf{e}_T^t$ have different shapes. All in all, $\mathbf{e}_s^t$ is actually similar to the supervision signal. By imitating $\mathbf{e}_s^t$, the target domain model can obtain cross-domain information. Note that as stated in \cite{Romero14}, the effect of middle layer distillation on the target domain model can also be regarded as a regularization, so in order to avoid over-regularization, the number and position of middle layer distillation need to be carefully selected. In this paper, for convenience, we select only the representation output by the last hidden layer as the intermediate representation by default.

\subsubsection{logit layer distillation} We further treat the logits output by the activation layer as soft labels, and then directly distill the extracted domain-invariant information through a high-temperature distillation method:
\begin{equation}
    L_{KL} = \frac{1}{|D^{Tar,t}|}\sum\limits_{i=1}^{|D^{Tar,t}|}\sigma(\frac{Z_{S_i}^t}{\tau})(log\sigma(\frac{Z_{S_i}^t}{\tau})-log\sigma(\frac{Z_{T_i}^t}{\tau})),
\end{equation}
where $\tau$ represents the temperature coefficient, and it can control the discrepancy between different distributions and precisely determine the difficulty level of KD \cite{Li23}. $\sigma(\cdot)$ is the softmax function. $Z_{S_i}^t$ and $Z_{T_i}^t$ are soft labels of source and target domain models respectively. Compared with the hard label that only has a value of 0 or 1, the soft label can provide a large amount of information contained in the negative label.

The overall loss of the KD in this paper is as follows:
\begin{equation}
    L_{KD} = \beta_1\sum\limits_{i=1}^{M}L_{MSE_i} +\beta_2L_{KL},
\label{eq9}
\end{equation}
where $M$ represents the number of the middle layer distillation. It is worth noting that the domain-invariant information migrator module allows a variety of source domain information storage approaches. For example, when you have a source domain model with a small number of parameters, you can save the model and output the information in the form of gradient truncation during the training phase. And when you have a domain with a small amount of data, you can also pre-store the information output by the source domain model.

\subsection{Optimization and Inference}
\label{Optimization and Inference}
In this section, we will first introduce the optimization of the recommendation task in the target domain, then introduce the overall optimization process and inference in detail.
\subsubsection{Train the Target domain Model}
Through the above operations, the target domain model has fully obtained the domain-invariant information. Next, we will mine the domain-specific information of the target domain through a model that is trained in the incremental mode. specifically, we train the period \textit{t}'s target domain model using only the latest incoming data. Our goal is to minimize the cross-entropy loss $L_{CE}$ as:
\begin{equation}
\hat{y} = f_T^t(x^{Tar,t}),
\label{eq.10}
\end{equation}
\begin{equation}
L_{CE} = -\frac{1}{|D^{Tar,t}|}\sum\limits_{i=1}^{|D^{Tar,t}|}y_ilog\hat{y}_i-(1-y_i)log(1-\hat{y}_i),
\label{eq11}
\end{equation}
where $\hat{y}$ represents the prediction of the target domain model.

\subsubsection{Overall Optimization}
 Finally, we will perform the overall loss optimization:
\begin{center}
\begin{equation}
L_{total}=
\begin{cases}
L_{adv1}& \text{step 1}\\
\lambda{L_{CE}} + \alpha{L_{adv2}} +\beta_1\sum\limits_{i=1}^{M}L_{MSE_i} +\beta_2L_{KL}& \text{step 2}
\end{cases}
\label{eq.12}
\end{equation}
\end{center}
where $\lambda$, $\alpha$, $\beta_1$ and $\beta_2$ is the hyper-parameters used to control the importance of different losses of DIIT, thereby ensuring that the target domain model preserves domain-specific information while accepting domain-invariant information.

It is worth noting that, as described in Section \ref{section Domain-invariant Information Extractors}, our optimization process is divided into two steps as shown in Eq. (\ref{eq.12}), which are performed in each epoch: first, update the discriminator parameter $\mathbf{\Theta}_{dis}^t$ by optimizing $L_{adv1}$. Next, update the remaining parameters by optimizing $L_{CE}$, $L_{adv2}$ and $L_{KD}$ simultaneously. In addition, the back-propagation of $L_{KD}$ cannot affect the parameters of the source domain models, thereby avoiding the loss of accuracy due to co-adaption of the source domain models and the target model \cite{Song22}. 

\subsubsection{Inference}
In the inference phase, Since KD can transfer information from the teacher model to the student model, and only the student model is needed for inference. Therefore, unlike models proposed by \cite{Zhang22,Liu23}, we only need to use the target domain model to make predictions for inference according to Eq. (\ref{eq.10}), which significantly improves the efficiency.

\section{Experiments}
\label{sec:Experiments}
In this section, we first introduce the experimental setup. Next, we design experiments to answer the following research questions:
\begin{itemize}
\item
\textbf{RQ1:} How does the effectiveness and efficiency of DIIT compared with state-of-the-art single-domain/cross-domain methods?
\end{itemize}
\begin{itemize}
\item
\textbf{RQ2:} As a pluggable unit, does DIIT perform well when used with different backbone models?
\end{itemize}
\begin{itemize}
\item
\textbf{RQ3:} How about the impact of each part on the overall model?
\end{itemize}
\begin{itemize}
\item
\textbf{RQ4:} What will be the impact when DIIT is introduced at different periods in incremental training?
\end{itemize}

\subsection{Experimental Setup}
\subsubsection{Datasets}
We conduct extensive experiments on one production dataset and two public datasets. These datasets have different magnitudes, ranging from hundreds of millions, tens of millions, and millions. The statistics of them are shown in Table \ref{tab:table1}.

\begin{table}
	\caption{Statistics of datasets}
	\centering
	\begin{tabular}{cc|ccc}
		\toprule
            \multicolumn{2}{c|}{\multirow{2}{*}{Dataset}} &\multicolumn{3}{c}{Domain} \\
		~ & ~ &A   &B   &C \\
            \midrule
            \multirow{2}{*}{Production} & Instances & 340M & 170M &230M \\
            & Percentage & 45.8\% & 22.8\% &31.4\% \\
            \midrule
            \multirow{2}{*}{Taobao} & Instances &8M &2M &8M \\
            & Percentage &45.1\% &10.3\% &44.6\% \\
            \midrule
            \multirow{2}{*}{KuaiRand} & Instances & 1M & 3M &400K \\
            & Percentage &22.4\% &69.1\% &8.5\% \\
		\bottomrule
	\end{tabular}
	\label{tab:table1}
\end{table}

\begin{itemize}           
\item
\textbf{Production.} Production is an advertising CTR prediction dataset that we collected in OPPO production system, which is one of the largest consumer electronics manufacturers in the world. Interaction logs from 2024-01-02 to 2024-01-08 are used for training and the next day for testing. Through discrete feature "\emph{domain ID}", the dataset can be divided into 10 domains. We select three of these domains for training and select the one with the smallest number of samples as the target domain. The production dataset is a large-scale dataset, in order to reduce the data size, we execute 10\% random negative sampling, and the ratio of positive and negative samples after it is about 1:5.
\end{itemize}

\begin{itemize}
\item
\textbf{Taobao.\footnote{https://tianchi.aliyun.com/dataset/dataDetail?dataId=56}} Taobao is a display advertising CTR prediction dataset provided by Alibaba. Interaction logs from 2017-05-06 to 2017-05-12 are used for training and the next day for testing. Following \cite{Zhang24} and \cite{Chen20}, we divide the dataset into different domains based on the discrete feature "\emph{city\_level}". We select three domains for training and select the one with the smallest number of samples as the target domain.
\end{itemize}

\begin{itemize}
\item
\textbf{KuaiRand \cite{Gao22}.} KuaiRand is a random recommendation dataset provided by Kuaishou. Interaction logs from 2022-04-16 to 2022-04-22 are used for training and the next day for testing. Through discrete feature "\emph{tab}", the dataset can be divided into 14 domains. We select three of these domains for training and select the one with the smallest number of samples as the target domain.
\end{itemize}

\begin{table*}[!t]
    \caption{Overall performance comparisons of DIIT with multiple single-domain and multi-domain models in the incremental mode on one production dataset and two public datasets. A higher AUC and a lower LogLoss represent a better performance.}
    \centering
    \vspace{2mm}
    \begin{tabular}{ccccccccccc}
    \toprule
    \multicolumn{2}{c}{\multirow{2}{*}{Model}} &\multicolumn{3}{c}{Production} &\multicolumn{3}{c}{Taobao} &\multicolumn{3}{c}{KuaiRand} \\
     ~ & ~ &AUC &LogLoss &Impr(AUC) &AUC &LogLoss &Impr(AUC) &AUC &LogLoss &Impr(AUC) \\
    \midrule
    \multirow{3}{*}{Single-domain} & DNN & 0.7455 & 0.0751 & +0.00\% & 0.5969 & 0.1928 & +0.00\% & 0.6648 & 0.6712 & +0.00\% \\
    ~ & DCN & 0.7472 & 0.0756 & +0.17\% & 0.5930 & 0.1930 & -0.39\% & 0.6663 & 0.6747 & +0.15\% \\
    ~ & W\&D & 0.7333 & 0.0761 & -1.22\% & 0.5834 & 0.1934 & -1.35\% & 0.6751 & 0.6695 & +1.03\% \\
    \midrule
    \multirow{4}{*}{Cross-domain} & DANN & 0.7429 & 0.0750 & -0.26\% & 0.5931 & 0.2011 & -0.38\% & 0.6735 & 0.6690 & +0.86\% \\
    ~ & HAMUR & 0.7454 & 0.0764 & -0.01\% & \textbf{0.6060} & 0.2030 & \textbf{+0.91}\% & 0.6807 & \textbf{0.6632} & +1.59\% \\
     ~ & CTNet & 0.7496 & 0.0745 & +0.41\% & 0.5991 & 0.1930 & +0.22\% & 0.6821 & 0.6640 & +1.73\% \\
    \midrule
    ~ & DIIT & \textbf{0.7526} & \textbf{0.0743} & \textbf{+0.71}\% & 0.5994 & \textbf{0.1923} & +0.25\% & \textbf{0.6826} & 0.6682 & \textbf{+1.78}\% \\
    \bottomrule
    \end{tabular}
    \label{tab:my_label2}
\end{table*}

\subsubsection{Evaluation Metrics and Implementation Details}
We apply AUC (Area Under Curve) and LogLoss (cross-entropy), which are commonly used in recommendation system as evaluation metrics. In general, a higher AUC or a lower LogLoss represents better performance. We use adam \cite{Kingma14} for optimization, the number of source domains is 2, the number of middle layer distillation is 1, the number of epochs is 1, the learning rate is 0.001, and the batch size is 4096. We fine-tuned the temperature coefficient within \{1,10,20,30,40,50\}, the coefficients of different losses within \{0,0.001,0.005,0.01,0.05,0.1,0.5,1,10,100\}. Note that we choose DNN as the backbone by default.

\subsubsection{Baselines}
To evaluate the performance of DIIT, we compare it with several single-domain/cross-domain methods and train them in the incremental mode.

\begin{itemize} 
\item
\textbf{Single-domain:} DNN \cite{Zhang16}: Deep Neural Network, a model of deep learning, which consists of MLPs. DCN \cite{Wang17}: Deep \& Cross Network, a DNN-based model, it introduces a novel cross-network that can learn certain bounded-degree feature interactions more effectively without significantly increasing the complexity. W\&D \cite{Cheng16}: Wide \& Deep Network, a model that combines the benefits of memorization and generalization by jointly training wide linear models and deep neural networks.
\item
\textbf{Cross-domain:} DANN \cite{Ganin16}: Domain-Adversarial Neural Network, a model that uses adversarial learning to align the representation distributions between the source domain and the target domain. HAMUR \cite{Li232}: a model that consists of a domain-specific adapter and a domain-shared hyper-network, and can be seamlessly integrated with various existing backbones as a plug-and-play component. CTNet \cite{Liu23}: Continual Transfer Network, a model that transfers knowledge from a time-evolving source domain to a time-evolving target domain. 
\end{itemize}

\begin{table}[t]
    \caption{Comparative experiment results on reference efficiency between CTNet and DIIT on the production dataset, The best results are shown in bold.}
    \centering
    \vspace{2mm}
    \begin{tabular}{ccc}
    \toprule
    Model & Time-consuming(s) & RelaImpr \\
    \midrule
    CTNet  & 191 & +0.00\% \\
    DIIT & \textbf{159} & \textbf{-16.75\%} \\
    \bottomrule
    \end{tabular}
    \label{tab:my_label3}
\end{table}

\begin{table*}[!t]
    \caption{Compatibility experiment results of DIIT with three different models as the backbone on the production dataset}
    \centering
    \vspace{2mm}
    \begin{tabular}{cccccccccc}
    \toprule
    \multirow{2}{*}{Method} &\multicolumn{3}{c}{Production} &\multicolumn{3}{c}{Taobao} &\multicolumn{3}{c}{KuaiRand} \\
    ~ & AUC & LogLoss & Impr(AUC) & AUC & LogLoss & Impr(AUC) & AUC & LogLoss & Impr(AUC) \\
    \midrule
    DNN & 0.7455 & 0.0751 & +0.00\% & 0.5969 & 0.1928 & +0.00\% & 0.6648 & 0.6712 & +0.00\% \\
    DNN+DIIT & 0.7526 & 0.0743 & +0.71\% & 0.5994 & 0.1923 & +0.25\% & 0.6826 & 0.6682 & +1.78\% \\
    \midrule
    DCN & 0.7472 & 0.0756 & +0.00\% & 0.5930 & 0.1930 & +0.00\% & 0.6663 & 0.6747 & +0.00\% \\
    DCN+DIIT & 0.7516 & 0.0743 & +0.44\% & 0.5993 & 0.1926 & +0.63\% & 0.6699 & 0.6743 & +0.36\% \\
    \midrule
    W\&D & 0.7333 & 0.0761 & +0.00\%  & 0.5834 & 0.1934 & +0.00\% & 0.6751 & 0.6695 & +0.00\% \\
    W\&D+DIIT & 0.7376 & 0.0758 & +0.43\% & 0.5866 & 0.1931 & +0.32\% & 0.6755 & 0.6697 & +0.04\% \\
    \bottomrule
    \end{tabular}
    \label{tab:my_label4}
\end{table*}

\begin{table}[t]
    \caption{Ablation experiment results of DIIT with DNN as the backbone on the production dataset, The best results are shown in bold.}
    \centering
    \vspace{2mm}
    \begin{tabular}{ccc}
    \toprule
    Model & AUC & Impr \\
    \midrule
    Base (DNN) & 0.7455 & +0.00\% \\
    DIIT (Only A) & 0.7508 & +0.53\% \\
    DIIT (Only C) & 0.7500 & +0.45\% \\
    DIIT (w/o Gating) & 0.7521 & +0.66\% \\
    DIIT (w/o Adversarial) & 0.7516 & +0.61\% \\
    DIIT (w/o Middle) & 0.7510 & +0.55\% \\
    DIIT (w/o Logit) & 0.7517 & +0.62\% \\
    DIIT & \textbf{0.7526} & \textbf{+0.71\%} \\
    \bottomrule
    \end{tabular}
    \label{tab:my_label5}
\end{table}

\subsection{Overall Performance (RQ1)}
In this section, we conduct extensive experiments to demonstrate the effectiveness and efficiency of DIIT. In addition to AUC and LogLoss, we also record the improvement of AUC. 

\subsubsection{effectiveness} To verify the effectiveness of DIIT in the industrial RS environment, we compare it with various advanced single-domain and cross-domain methods. The experimental results are shown in Table \ref{tab:my_label2}, and our observations are as follows.
\begin{itemize}           
\item
DIIT's performance is advantageous in most cases, especially when compared with single-domain methods, which demonstrates the importance of transferring domain-invariant information across domains. It is worth noting that HAUAMR performs well on the Taobao dataset, its AUC is significantly higher than various state-of-the-art models including DIIT. However, HAMUR is based on an ideal static assumption that users' interests do not change much in a short period, which is impractical in the industrial RS environment.
\item
We particularly compare DIIT with CTNet, and find that CTNet's performance is slightly worse than DIIT in most cases. The reasons for this are: 1) through two extractors, DIIT can extract domain-invariant information more efficiently. 2) CTNet is designed for the case where the source domain number is 1, which limits its ability to obtain more information across domains.
\end{itemize}

\subsubsection{efficiency} To verify the efficiency of DIIT in the industrial RS environment, we compared it with KEEP \cite{Zhang22} and CTNet \cite{Liu23}, which have similar research directions to this paper. Among them, KEEP is a two-stage framework which need additional computation and storage resources. In contrast, DIIT directly extracts and transfers the domain-invariant information from multiple source domain models that train in the incremental mode independently.

CTNet is a more lightweight method that also does not need additional computation and storage resources. However, it needs to feed samples to both the source and the target domain model, whether in the training or inference phases. In contrast, DIIT only needs to keep the target domain model in the inference phase. As shown in Table \ref{tab:my_label3}, on the sampling test dataset in the period \textit{t}, DIIT's inference time-consuming is reduced by about 16.75\%. As mentioned above, CTNet only considers the situation of one source domain, while DIIT can be easily applied to the situation of multiple source domains, and as the number of source domains increases, the gap of inference time-consuming will become larger.

\subsection{Compatibility Experiment (RQ2)}
As a pluggable component, it is necessary to verify the validity of DIIT with different backbones. In this section, we choose DNN, DCN, and W\&D as the backbone respectively. The experimental results are shown in Table \ref{tab:my_label4}. Similar to Table \ref{tab:my_label2}, we also bold the best results and omit the improvement of LogLoss while marking the AUC improvement. We can found that no matter which network is used as the backbone, the proposed DIIT can bring improvements. It proves that DIIT can not only well help the target domain model obtain the domain-invariant information across domains, but also has good practicability.

\subsection{Ablation Experiment (RQ3)}
To verify the effectiveness of each module of DIIT, we design extensive ablation experiments by removing different modules. Specifically, we considered the following cases: 1) Use only one domain as the source domain (Only A or Only C); 2) Remove the domain-invariant information extractor at the domain level and use summation to aggregate multiple source domain model representations (w/o Gating); 3) Remove the domain-invariant information extractor at the representation level (w/o Adversarial); 4) Use only the middle distillation or the logit layer distillation for KD (w/o Middle \& w/o Logit). Note that we also recorded the experimental results of DNN (Base) for intuitive comparison. The experimental results are shown in Table \ref{tab:my_label3}, best results are shown in bold, and in addition to recording AUC, we also marked the improvement of AUC. The improvement trend of LogLoss is similar to AUC, due to limited space, we omit it in Table \ref{tab:my_label5}. Our observations are as follows:

\begin{figure}[!t]
    \centering
    \begin{tabular}{@{\extracolsep{\fill}}c@{}c@{\extracolsep{\fill}}}
            \includegraphics[scale=0.21]{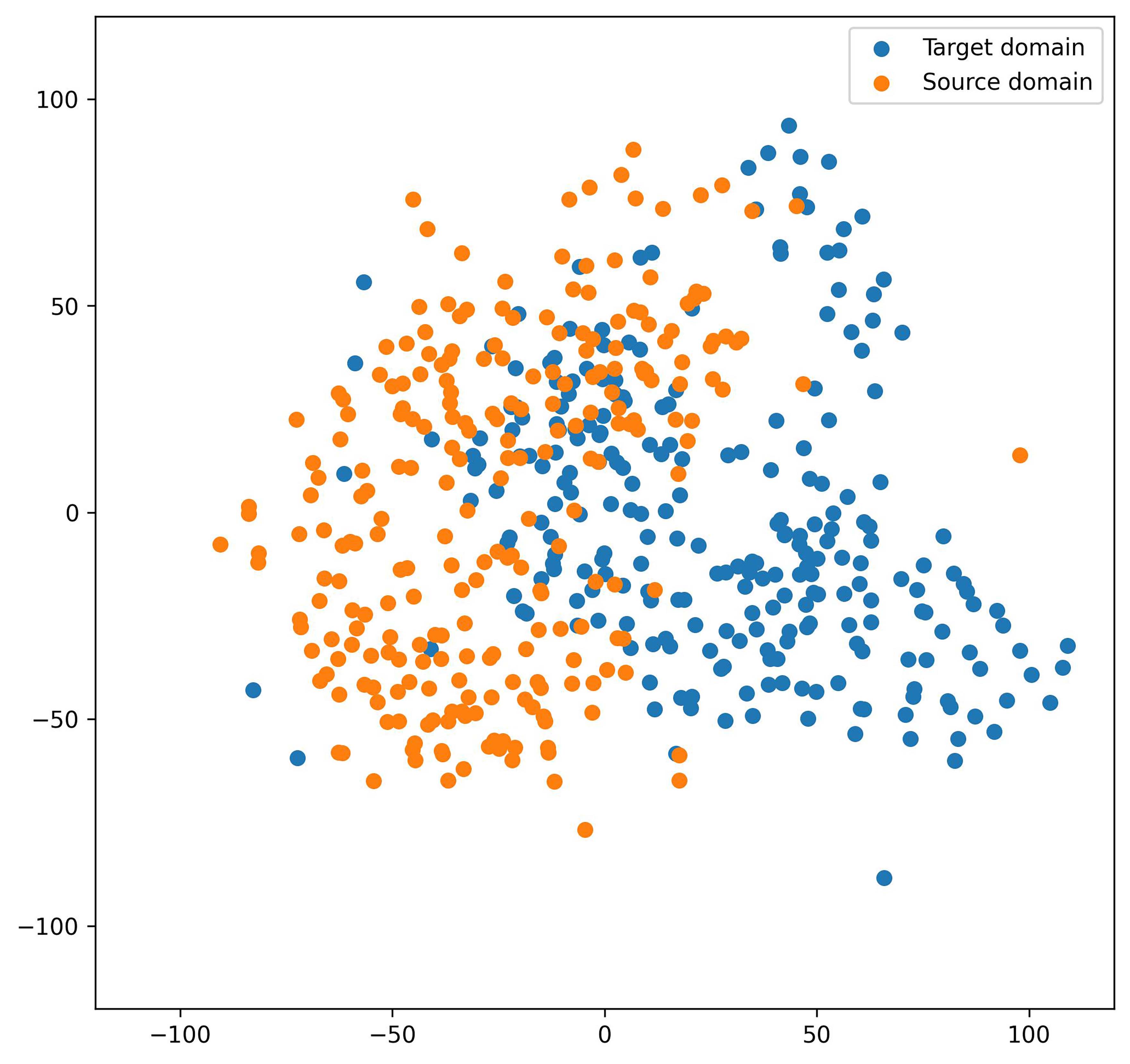} &
            \includegraphics[scale=0.21]{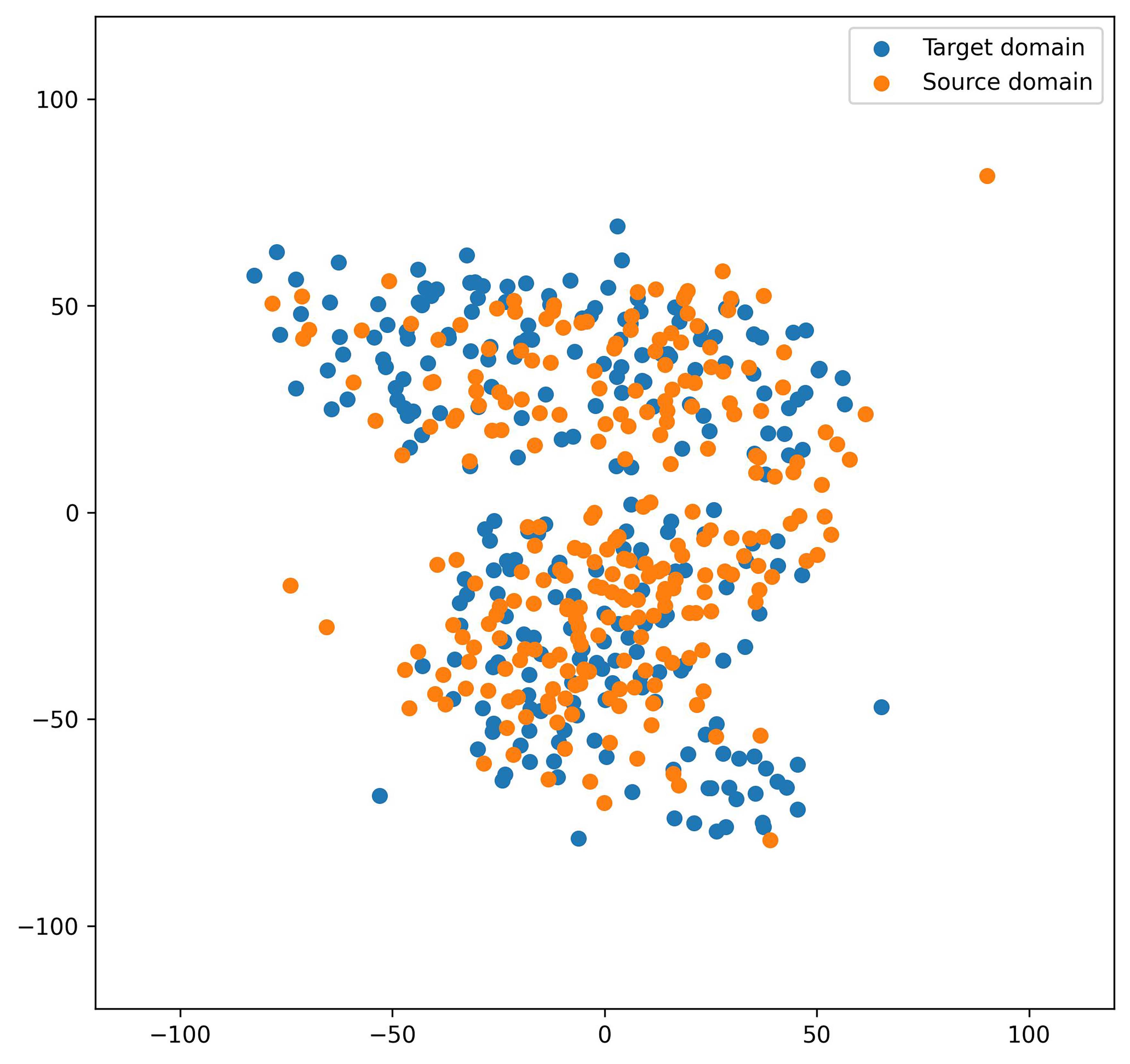}\\
            (a) Before & (b) After \\
    \end{tabular}
    \caption{The t-SNE visualization of representations before and after the domain-invariant information extractors on the production dataset.}
    \Description{Figure 3 visualizes the representations before and after the domain-invariant information extractors to show that the extractors can confuse the distribution of representations from the source and target domains, thereby ultimately extracting the domain-invariant information.}
    \label{fig:image_with_tabletsne}
 \end{figure}

\begin{itemize}           
\item
Comparing DIIT, DIIT (Only A) and DIIT (Only C), we can find that using more source domains achieves better results, which shows that different source domains can complement each other, thereby providing richer information to the target domain model.
\end{itemize}

\begin{itemize}           
\item
Comparing DIIT and DIIT (w/o Gating), we can find that using the target domain model to guide the aggregation of multiple source domain models through a gating network achieves better results, which proves the necessity of the domain-invariant information extractor at the domain level.
\end{itemize}

\begin{itemize}           
\item
Comparing DIIT and DIIT (w/o Adversarial), we can find that by using adversarial learning to align the representation distributions output by the source domain models and the target domain model, the domain-invariant information in the source domain is transferred to the target domain model in a fine-grained manner, and effectively improve the effectiveness of the target domain model.
\end{itemize}

\begin{itemize}           
\item
Comparing DIIT, DIIT (w/o Middle) and DIIT (w/o Logit), we can find that distilling information from different spots of the source domain models to the target domain model achieves better results, which proves the necessity of the multi-spot KD.
\end{itemize}

\begin{itemize}           
\item
Finally, no matter which version of DIIT is compared with Base, there is a significant improvement, which proves the importance of applying the CDR methods in the industrial RS environment.
\end{itemize}

\textbf{Visualization.} To provide a more comprehensive insight of DIIT, we visualize representations before and after the domain-invariant information extractors by t-SNE \cite{Maaten08}. As shown in Figure \ref{fig:image_with_tabletsne}, representations of the source and target domains become more inseparable, indicating that their distributions are aligned while domain-invariant information is extracted.

\begin{figure}[!t]
    \centering
    \begin{tabular}{@{\extracolsep{\fill}}c@{}c@{\extracolsep{\fill}}}
            \includegraphics[scale=0.14]{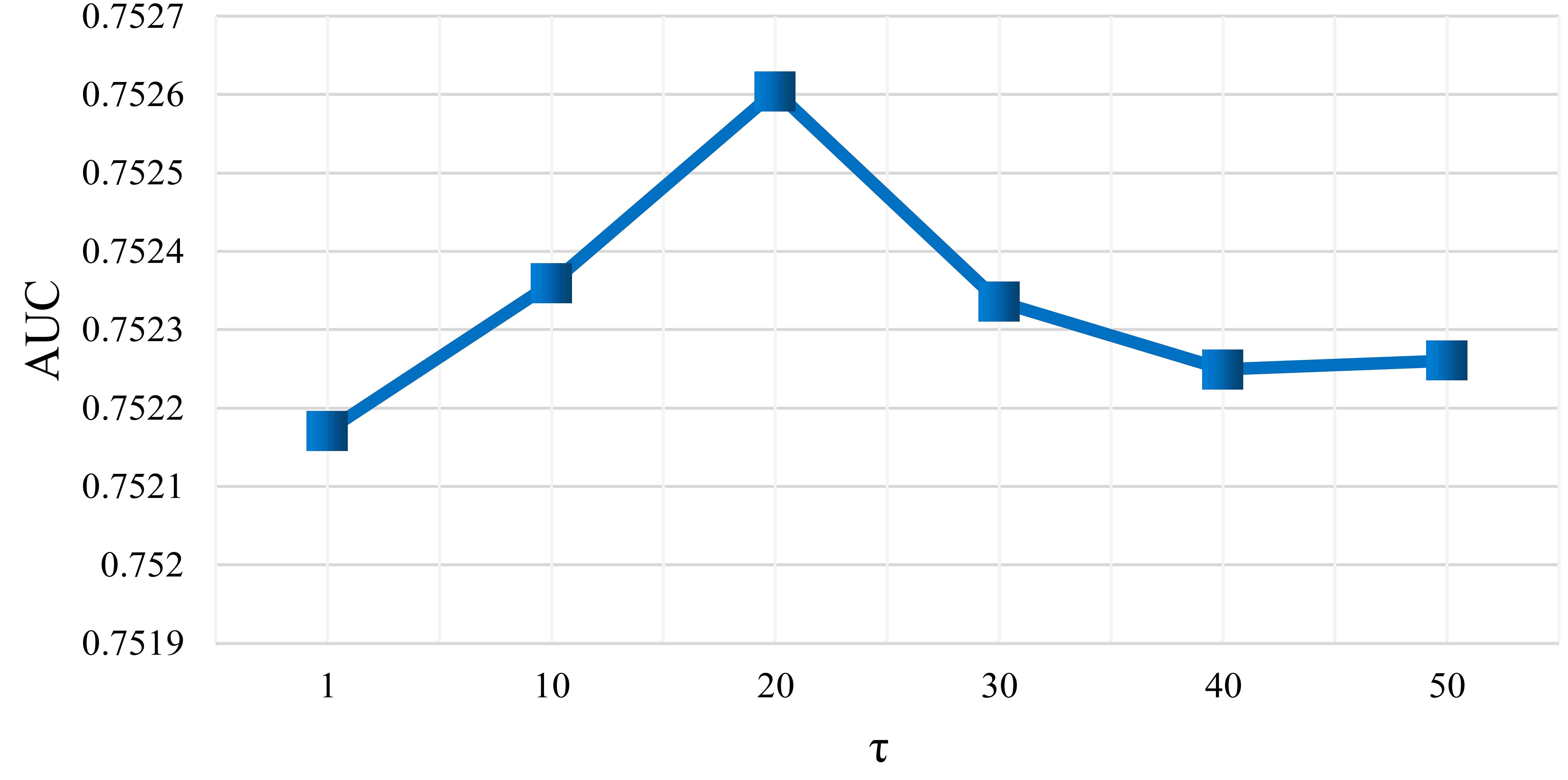} &
            \includegraphics[scale=0.14]{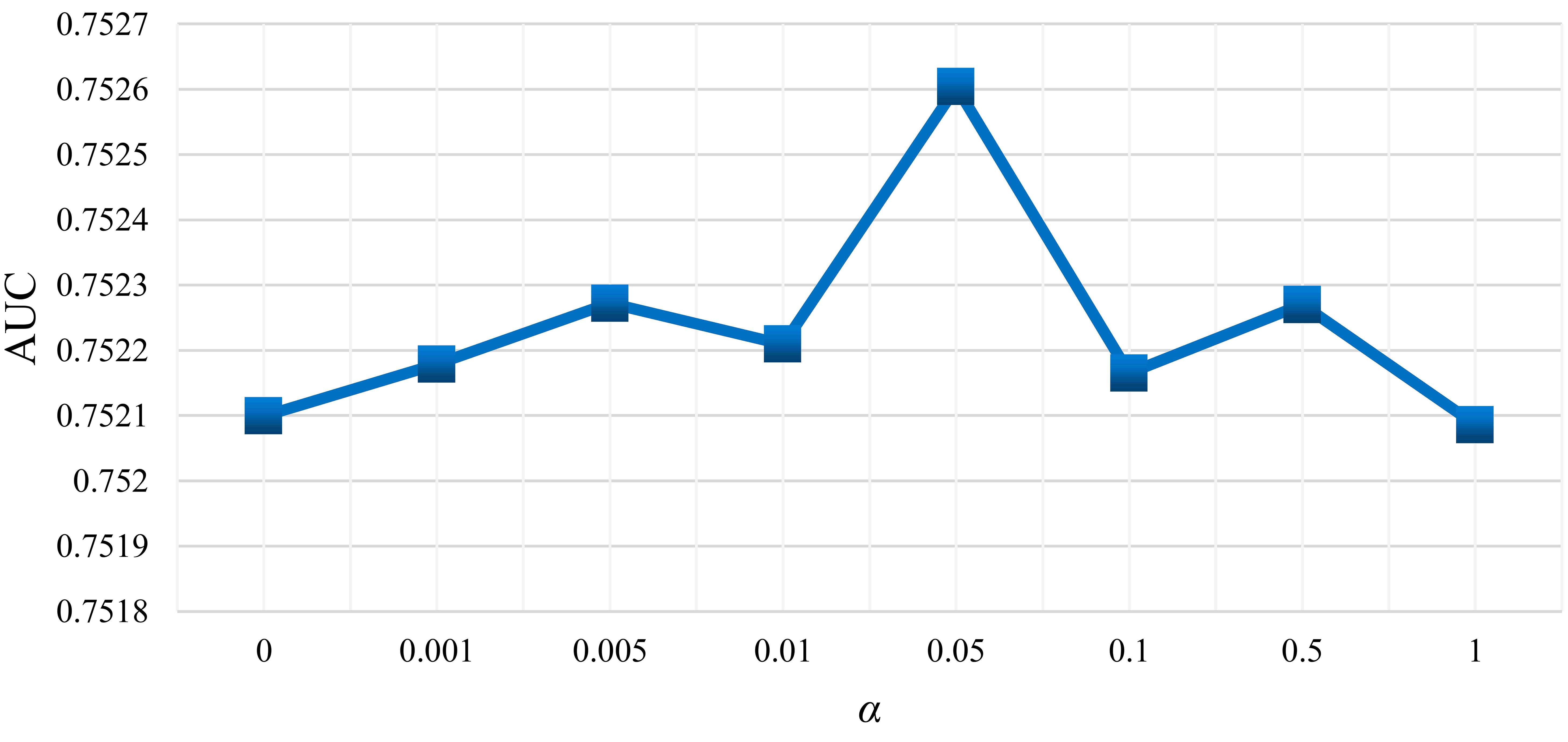}\\
            (a) $\tau$ & (b) $\alpha$\\
            \includegraphics[scale=0.14]{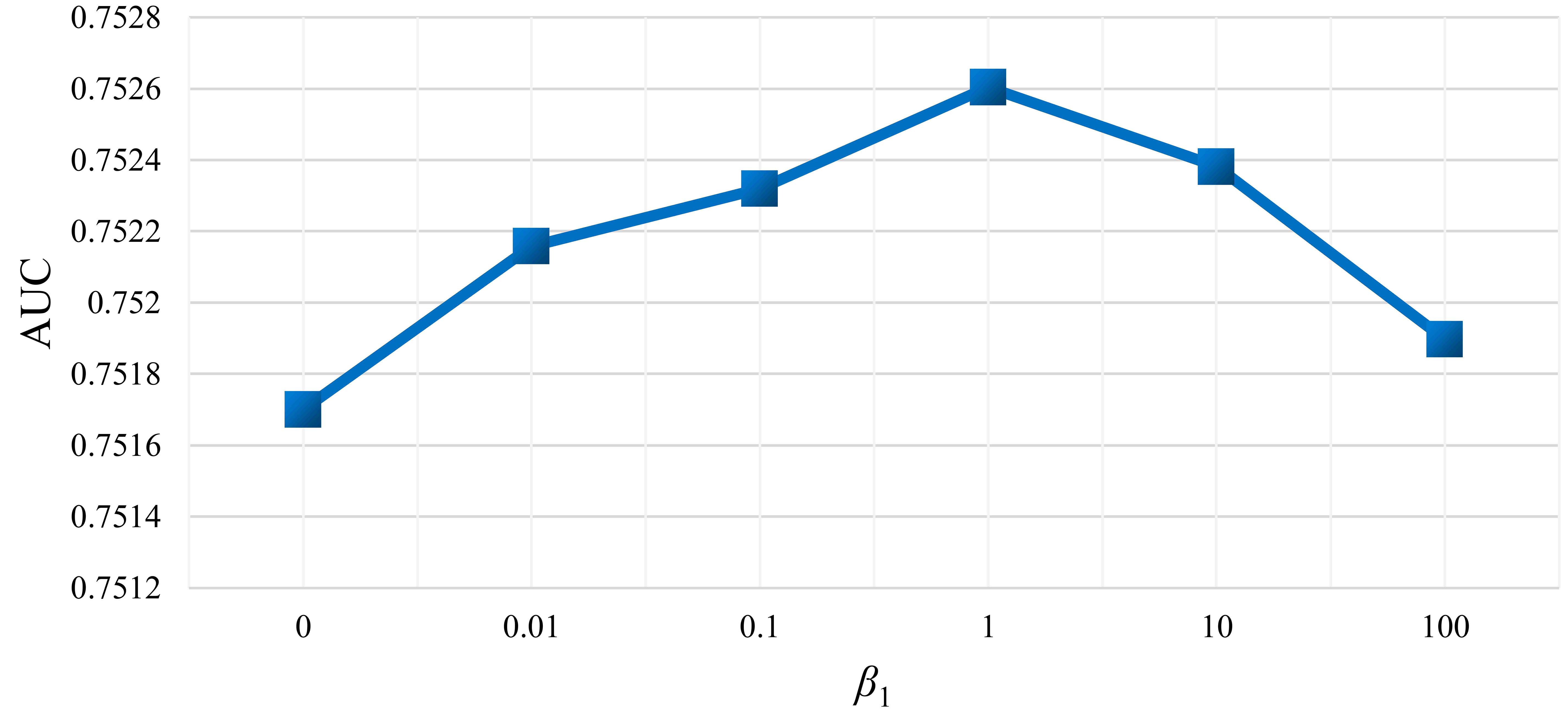} & 
            \includegraphics[scale=0.14]{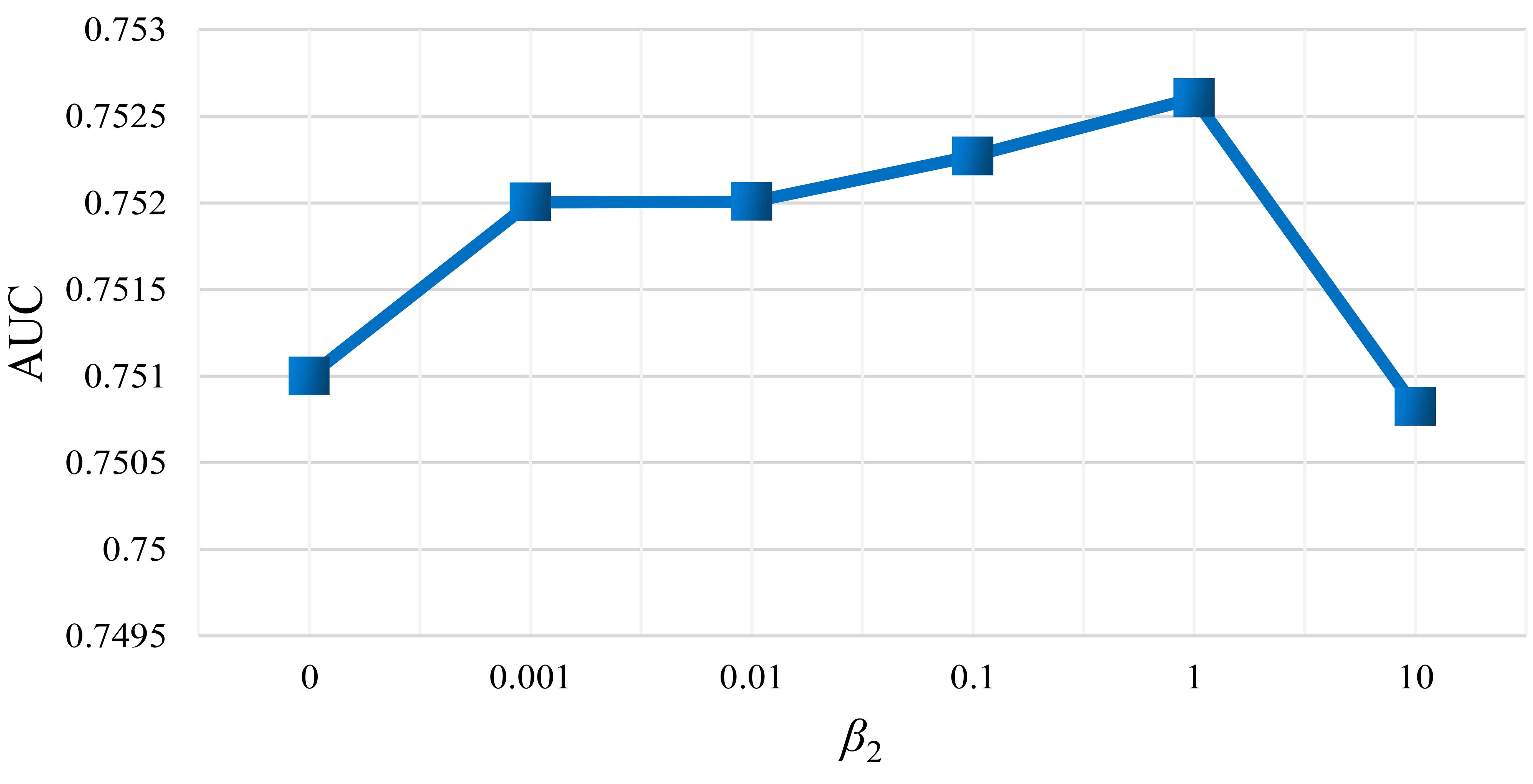}\\
            (c) $\beta_1$ & (d) $\beta_2$\\
    \end{tabular}
    \caption{Hyper-parameter experiment results of DIIT with DNN as the backbone on the production dataset.}
    \Description{Figure 4 shows the hyper-parameter experimental results on the coefficients of the temperature coefficient, the adversarial loss, the middle layer distillation and the logit layer distillation loss.}
    \label{fig:image_with_tablehp}
 \end{figure}

\subsection{Hyper-parameter Experiment (RQ3)}
In this section, we further design hyper-parameter experiments to observe the impact of different modules in DIIT. As mentioned before, DIIT involves multiple hyper-parameters, most of which have been introduced in previous works. Therefore, we focus on the coefficient $\tau$ of the distillation temperature, the coefficient $\alpha$ of the adversarial loss $L_{adv2}$, the coefficient $\beta_1$ of the middle layer distillation loss $L_{MSE}$, and the coefficient $\beta_2$ of the logit layer distillation loss $L_{KL}$. The experimental results are shown in Figure \ref{fig:image_with_tablehp}, and our observations are as follows:

\begin{itemize}           
\item
$\tau$: The temperature coefficient indicates how much the model pays attention to negative labels during the distillation. We find that a reasonable temperature can not only maximize the extraction of domain-invariant information from the source domain models, but also inhibit the transfer of source domains' domain-specific information.
\item
$\alpha$, $\beta_1$ and $\beta_2$: These coefficients represent the importance of different losses in DIIT respectively. We find that they have their own best performance ranges. By balancing them, the effectiveness of DIIT can be greatly improved. It is worth noting that when the coefficient of $L_{adv2}$ is 0, the remaining losses is actually used as the other side of the adversarial network to align the source domains and the target domain. However, by observing Figure \ref{fig:image_with_tablehp} and DIIT (w/o Adversarial) in Table \ref{tab:my_label5} simultaneously, we can find that both $L_{adv1}$ and $L_{adv2}$ have positive effects, and retaining both is the best choice.
\end{itemize}

\subsection{Exploratory Experiment (RQ4)}
\begin{figure}[!t]
  \centering
  \includegraphics[scale=0.9]{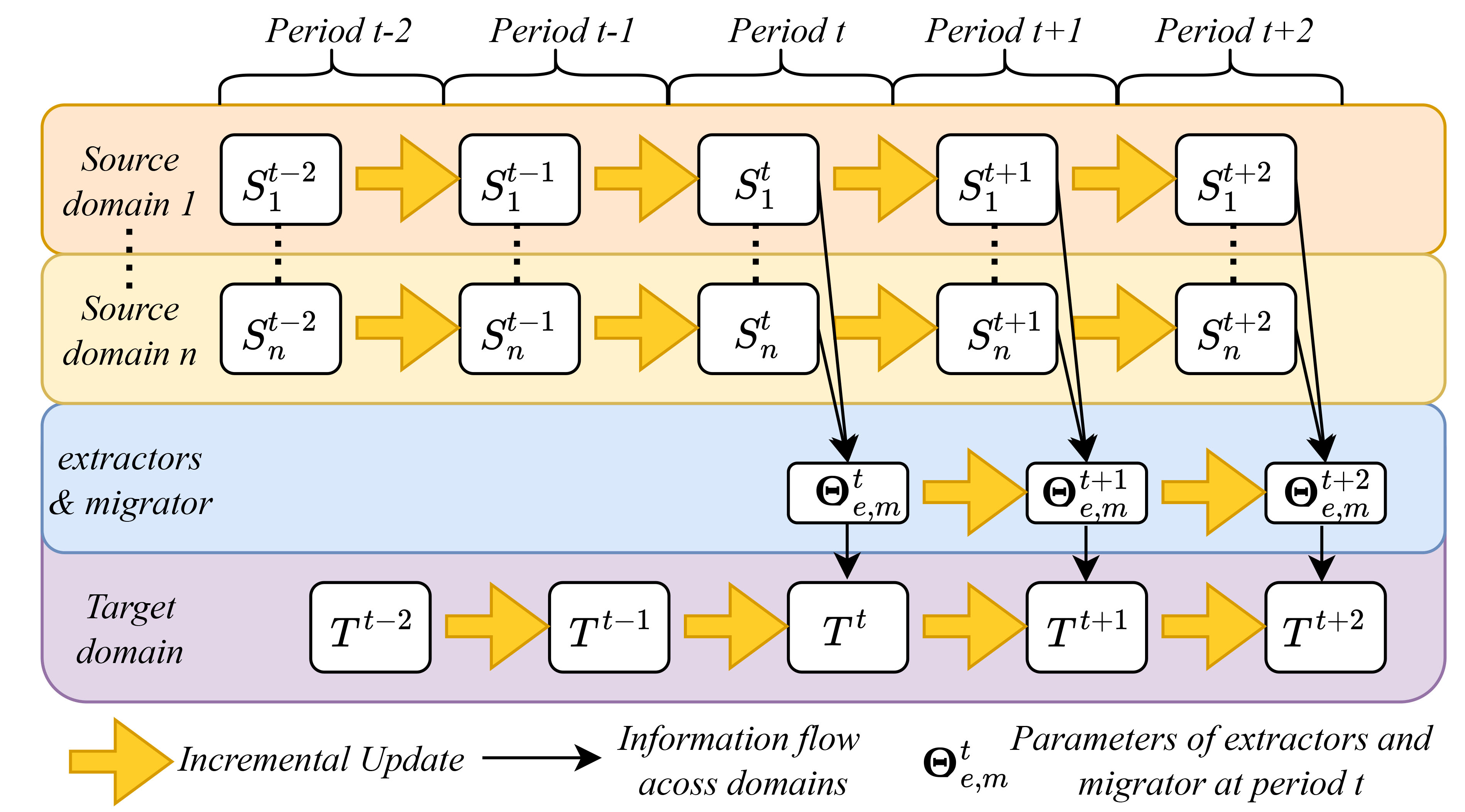}
  \caption{An illustration of how DIIT works in the industrial RS environment.}
  \Description{Figure 5 shows an example of how to plug DIIT in the industrial RS environment.}
  \label{fig:model_architecture3}
\end{figure}

\begin{figure}[!t]
    \centering
    \begin{tabular}{@{\extracolsep{\fill}}c@{}c@{\extracolsep{\fill}}}
            \includegraphics[scale=0.28]{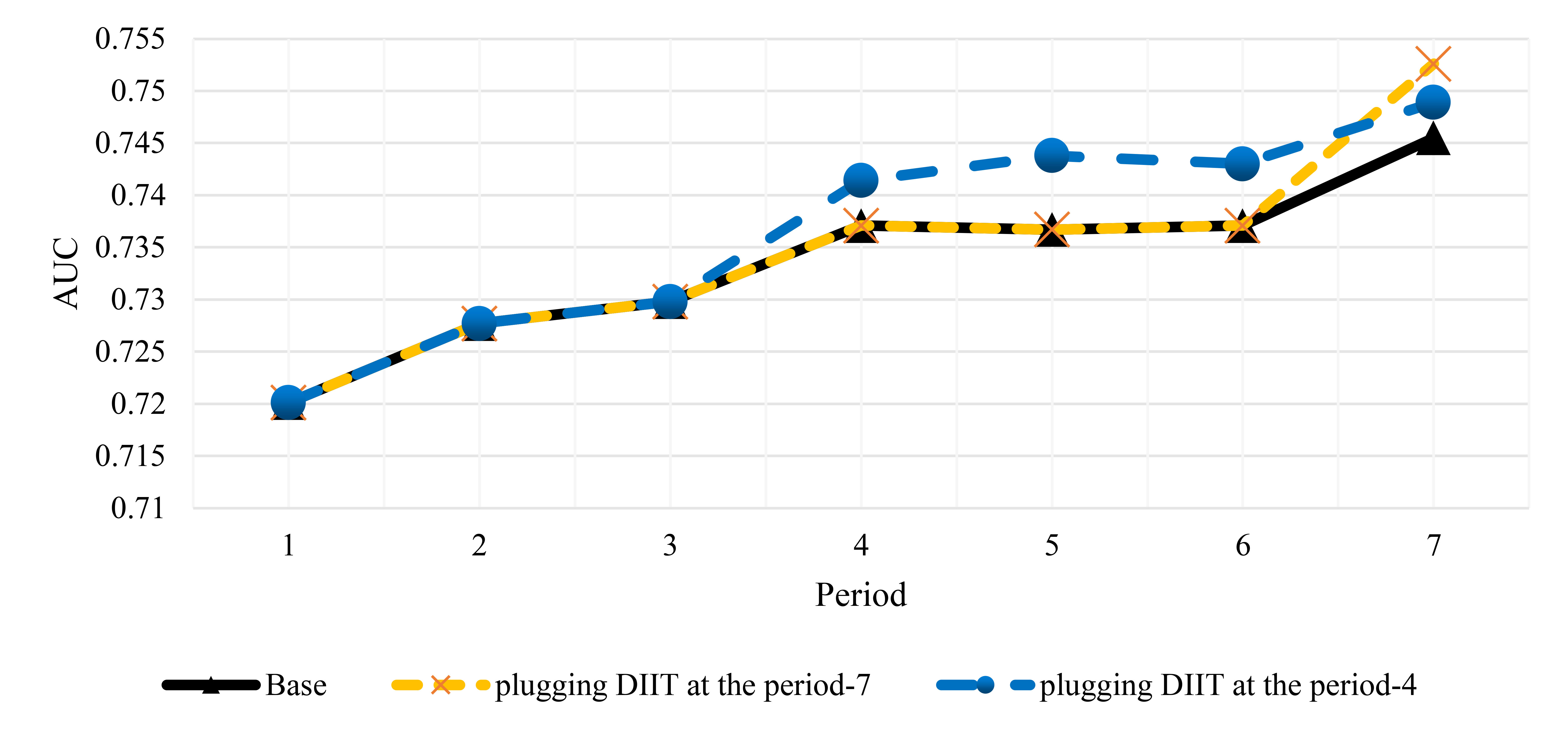} &
            \includegraphics[scale=0.28]{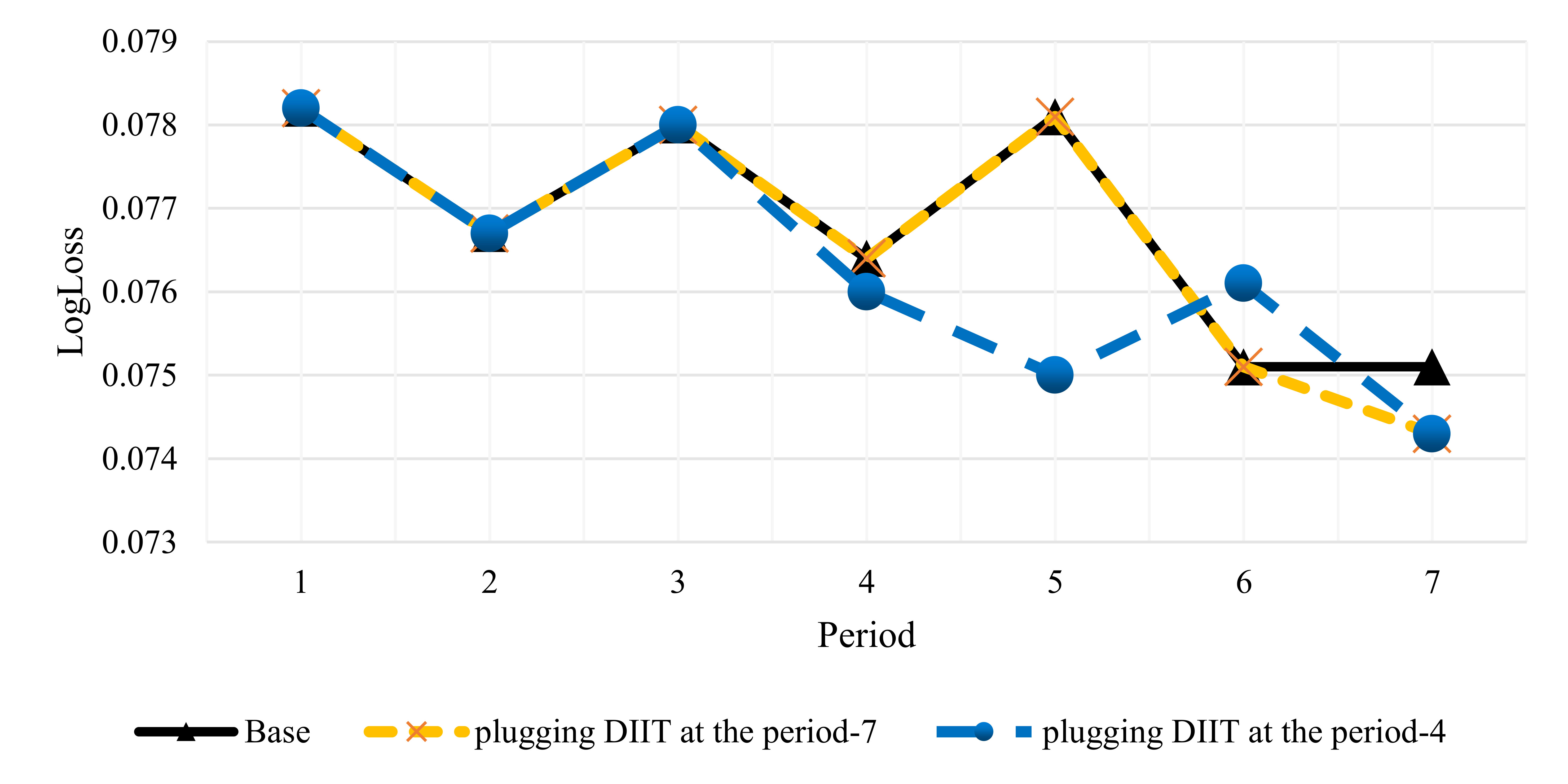}\\
            (a) AUC & (b) LogLoss\\
    \end{tabular}
    \caption{Exploratory experiment results of DIIT with DNN as the backbone on the production dataset. }
    \Description{Figure 6 shows the impact of plugging DIIT at different periods of incremental training.}
    \label{fig:image_with_table4}
 \end{figure}

As shown in Figure \ref{fig:model_architecture3}, we further explored more situations that may be faced in the industrial RS environment. As mentioned above, we consider a cross-domain recommendation task that contains multiple domains while each domain maintains its own model. In addition, we are curious about the impact of plugging the proposed DIIT in different periods under incremental learning (e.g. plugging DIIT in the period \textit{t} as shown in Figure \ref{fig:model_architecture3}). Therefore, we conducted extensive experiments to compare the effects of base and DIIT that is plugged in period 4 and period 7 of a 7-period incremental training. The experimental results are shown in Figure \ref{fig:image_with_table4}, we find that plugging DIIT into the base model in different periods can bring positive effects in most cases. Moreover, comparing the curves of period 4 and period 7, it can be found that earlier plugging does not mean a better result. This may be caused by too much domain-invariant information from previous source domains flooding the domain-specific information of the latest target domain, which indicates that we can expect to observe the benefits after plugging DIIT in a short time.

\section{Conclusion}
In this paper, we propose DIIT, an end-to-end domain-invariant information transfer method for industrial cross-domain recommendation. We first simulated the industrial recommendation systems (RS) environment, where multiple domains maintain respective models and train them in the incremental mode. Next, in order to improve the effectiveness and efficiency of cross-domain recommendation in the industrial RS environment, we design two extractors to extract domain-invariant information at the domain level and the representation level respectively, and then design a migrator to transfer them to the target domain model. DIIT is plug-and-play and can be integrated with various methods. We further conduct extensive experiments on three datasets of different magnitudes, including one production dataset and two public datasets, to demonstrate the effectiveness and efficiency of DIIT. Future work will focus on the application of the cross-domain recommendation method in more complex environments, such as when the number of source or target domains is dynamically changing.

\bibliographystyle{ACM-Reference-Format}
\balance
\bibliography{Reference}

\appendix
\section{Appendix}

\subsection{Training Process of DIIT}
we summarize the detailed training process in the period \textit{t} of DIIT in \textbf{Algorithm \ref{algorithm1}}.

\begin{algorithm}[b]
\SetKwData{Left}{left}\SetKwData{This}{this}\SetKwData{Up}{up}
  \SetKwFunction{Union}{Union}\SetKwFunction{FindCompress}{FindCompress}
  \SetKwInOut{Input}{input}\SetKwInOut{Output}{output}
\caption{Training Process in the period \textit{t} of DIIT}\label{algorithm1}
\Input{Number of source domains $N$, the latest source domain $S_n^t$, the latest target domain $T^{t-1}$, the latest trainable parameters $\Theta_{gate}^{t-1}$, $\Theta_{mapper}^{t-1}$, $\Theta_{dis}^{t-1}$, the target domain sample $x^{Tar,t} \in D^{Tar,t}$, the mix domain sample $x^{Mix,t} \in D^{Mix,t}$.}
\Output{The target domain model $T^{\,t}$.}
\tcp{Warm Start}
\uIf {plugging DIIT for the first time} {
    Initialize $T^{\,t}$ based on $T^{\,t-1}$, Initialize Parameters $\Theta_{gate}$, $\Theta_{mapper}^{t}$, $\Theta_{dis}^{t}$ randomly\;
} \Else{
    Initialize $T^{\,t}$,$\Theta_{gate}^{t}$, $\Theta_{mapper}^{t}$, $\Theta_{dis}^{t}$ based on $T^{\,t-1}$,$\Theta_{gate}^{t-1}$, $\Theta_{mapper}^{t-1}$, $\Theta_{dis}^{t-1}$\;
}
\For{each epoch}{\For{each mini-batch}{
\tcp{Domain-invariant Information Extractors}
Get $\mathbf{e}_{s_n}^t, \mathbf{Z}_{s_n}^t\leftarrow S^t_n(x^{Tar,t})$,  $\mathbf{e}_{adv_{s_n}}^t\leftarrow S^t_n(x^{Mix,t})$\;
Get $\mathbf{e}_{S}^t$, $\mathbf{e}_{adv}^t$ and $\mathbf{Z_S}^t$ by Eq. (\ref{eq1})\;
Get $\mathbf{e}_{adv}^t$ by Eq. (\ref{eq3})\; 
Calculate $L_{adv1}$ by Eq. (\ref{eq5})\; 
Update $\Theta_{dis}^t$ by using Adam\; 
Calculate $L_{adv2}$ by Eq. (\ref{eq6})\; 
Get $\mathbf{e}_{S}^t$ by Eq. (\ref{eq3})\; 
\tcp{Domain-invariant Information Migrator}
Calculate $L_{KD}$ by Eq. (\ref{eq9})\;
\tcp{Train the Target Model}
Calculate  $L_{CE}$ by Eq. (\ref{eq11})\;
\tcp{Overall Optimization}
Calculate the total loss $L_{total}$ by Eq. (\ref{eq.12})\; 
Update $\Theta_{target}^t$, $\Theta_{gate}^t$, and $\Theta_{mapper}^t$ by using Adam}}
\Return $T^{t}$
\end{algorithm}

\end{document}